\documentclass[12pt,preprint]{aastex}

\begin{document}

\title{Evidence for Tidal Effects in the Stellar Dynamics of the Large Magellanic Cloud}

\author{Knut A.G. Olsen\affil{National Optical Astronomy Observatory,
CTIO, Casilla 603, La Serena, Chile\\ kolsen@ctio.noao.edu}}

\author{Philip Massey\affil{Lowell Observatory, 1400 W. Mars Hill Road, Flagstaff, AZ 86001 \\ Phil.Massey@lowell.edu}}

\begin{abstract}
We present a combined analysis of the kinematics of the Large
Magellanic Cloud through its \ion{H}{1} gas, carbon stars, and red
supergiant stars.  After correcting the line-of-sight velocities for
the recent accurate measurement of the LMC's space motion, we find
that each kinematic tracer clearly defines a flat rotation curve with
similar shape but different amplitude for each tracer: 61 km s$^{-1}$
for the carbon stars, 80 km s$^{-1}$ for the \ion{H}{1}, and 107 km
s$^{-1}$ for the red supergiants.  Previously identified tidal
\ion{H}{1} features are seen to harbor numerous carbon stars, with the
tidally disturbed stars comprising 7--15\% of the total sample.  This
discovery implies that we cannot depend on the carbon star sample
alone to construct a reliable model of the LMC's gravitational
potential.  We also find red supergiants with peculiar kinematics, but
their association with tidal features is unclear, and may instead be
interacting with supergiant \ion{H}{1} shell SGS4.  In addition,
although the local velocity dispersion of the red supergiants is
small, $\sim$8 km s$^{-1}$, their velocity dispersion about the carbon
star rotation solution is 17 km s$^{-1}$, equal to the velocity
dispersion of the carbon stars themselves.  We thus appear to be
witnessing the tidal heating of the LMC's stellar disk.
\end{abstract}

\section{Introduction}
The disturbances inflicted on the \ion{H}{1} gas of the Large
Magellanic Cloud by the Small Magellanic Cloud and the Milky Way
Galaxy are well documented.  The most famous feature is the Magellanic
Stream (Wannier \& Wrixon 1972, Mathewson et al.\ 1974), to which the
SMC and LMC both contribute (Staveley-Smith et al.\ 2003), and whose
Leading Arm (Putman et al.\ 1998) indicates a tidal origin.  The
large-scale \ion{H}{1} of the LMC is further marked by four spiral
arm-like features, whose connection with the Stream, the Magellanic
Bridge, and the Leading Arm suggests that they were also formed by
tidal disturbances (Staveley-Smith et al.\ 2003).  The LMC's stellar
population also shows evidence of these tidal interactions.
The dominant
stellar component is a disk, with the most recent measurements of its
inclination giving $i\sim35^\circ$ and position angle of the line of
nodes $\Theta\sim130-150^\circ$ (van der Marel \& Cioni 2001, Olsen \&
Salyk 2002, Nikolaev et al.\ 2004).  Once corrected for the effect of
the inclination, the disk appears to be intrinsically elongated in the
direction of the Milky Way (van der Marel 2001).  Alves \& Nelson
(2000) demonstrated that the disk is flared, while both Olsen \& Salyk
(2002) and Nikolaev et al.\ (2004) found evidence that it is
warped.  Kunkel et al.\ (1997) suggested that the LMC harbors a polar
ring on the basis of a number of carbon stars with peculiar
velocities, while Graff et al.\ (2000) identified a possible
kinematically distinct population of carbon
stars, suggesting that the population lies outside the LMC.  To date,
however, searches for tidally stripped LMC stars clearly associated
with its tidal \ion{H}{1} features have proved unsuccessful.  These
searches have naturally focussed on areas far away from the LMC, where
the tidal \ion{H}{1} features are prominent.  In this Letter, we use
kinematic evidence to demonstrate that a significant fraction of the
stars in the inner regions and near periphery of the LMC may, in fact,
lie in the tidal arms.

\section{Observational Data}
We used three sets of observations to compare the kinematics of the LMC's \ion{H}{1} gas and stellar populations.  The \ion{H}{1} observations are those from Kim et al.\
(2003), which derive from the Australia Telescope Compact Array (ATCA)
survey of Kim et al.\ (1998) combined with the Parkes multibeam survey
of Staveley-Smith et al.\ (2003).
From the combined \ion{H}{1} datacube, we calculated the \ion{H}{1} emission-weighted mean velocity at every spatial pixel, which we then used for the analysis in this paper.  To probe the intermediate-age and older stellar populations, we used the carbon star velocity samples obtained by Kunkel et al.\ (1997) and by Hardy, Schommer, \& Suntzeff (2006), which have been extensively analyzed by Graff et al.\ (2000), Alves \& Nelson (2000), and van der Marel et al.\ (2002; hereafter vdM02).  Our young star kinematic sample consists of the LMC red supergiants (RSGs) described in Massey \& Olsen (2003; Paper 1) supplemented by the RSG catalog of Pr\'{e}vot et al. (1985), whose kinematics were discussed by Pr\'{e}vot et al. (1989).  The Paper I RSG velocities have measured velocities good to $<$1 km s$^{-1}$, whereas the typical errors for the Pr\'{e}vot et al. (1985) catalog are $1.5 - 2.0$ km s$^{-1}$.

\section{Analysis}
As noted by many previous authors (Meatheringham et al.\ 1988, Schommer et al.\ 1992, vdM02), the interpretation of line-of-sight velocities of LMC sources must account for the contribution of the LMC's space motion projected into the line of sight.  Because the angle subtended on the sky by the LMC is large, the magnitude of the projected space motion vector depends on location.  
As pointed out by vdM02, an added complication is that the LMC may be undergoing precession and nutation, which would appear as a velocity gradient in the direction perpendicular to the kinematic line of nodes.  
The recent accurate measurement of the LMC's proper motion by Kallivayalil et al.\ (2006) is thus very important for interpreting the LMC's internal kinematics.  As Kallivayalil et al.\ note, the new proper motion measurement significantly improves the constraint as set by the carbon stars on the rate of change of the LMC disk's inclination d$i$/d$t$, such that it is now consistent with zero, and places the carbon star rotation velocity at $V_0\sim60$ km s$^{-1}$.  We confirmed these values by fitting the carbon star velocities to equation 24 of vdM02, where we solved for the location of the center of mass, the systemic velocity $v_{\rm sys}$, d$i$/d$t$, the position angle of the line of nodes $\Theta$, and the LMC's rotation curve using the parameterization of equation 36 of vdM02.  
Following the procedure outlined in vdM02, we obtained $\alpha_{\rm CM}=5^{\rm h}30\fm2$, $\delta_{\rm CM}=-69\fdg97$, $v_{\rm sys}=263.0$ km s$^{-1}$, d$i$/d$t=0.001$ mas yr$^{-1}$, $\Theta=131\fdg0$, rotation curve amplitude $V_0=61.1$ km s$^{-1}$, rotation curve scale length $R_0/d_0=0.041$ ($d_0\equiv50.1$ kpc), and a rotation curve shape parameter of $\eta=3.00$ with a velocity dispersion $\sigma=19.2$ km s$^{-1}$ around the fit; these values are all in agreement with those of vdM02, within their errors, after accounting for the effect of the updated proper motion value of Kallivayalil et al.\ (2006).  Because our aim was simply to show that we are able to reproduce vdM02's results, we did not perform the Monte Carlo simulations needed to calculate errors in the parameters, as done by vdM02.

The transverse motion of the LMC dominates the line-of-sight velocities, with the LMC's internal motions introducing only a distortion on the velocity gradient produced by the space motion.  
In order to directly compare the internal kinematics of the LMC's \ion{H}{1}, red supergiants, and carbon stars, we subtracted the component of the carbon star kinematic model containing the space motion 
from the velocity sets.  We further divided these velocity residuals by the factor $g(\rho,\Phi)=f(\rho,\Phi)\sin i\cos(\Phi-\Theta)$, where we set $i=34\fdg7$ and $\Theta=131\fdg0$ and where all other terms are defined in equations 24 and 25 of vdM02: $\rho$ is the angular distance from the center of mass, $\Phi$ is the position angle on the sky with respect to the center of mass, and the function $f(\rho,\Phi)$ describes the projection of the in-plane radius $R^\prime$ onto the sky.  Under the assumption that the gas and stars all lie in the LMC disk and are on purely circular orbits, this last operation yields their rotation curves as a function of the in-plane radius $R^\prime$.  Any significant out-of-plane motions should appear to have velocities substantially different from that of the rotation curve in such a presentation.
We excluded regions with $g(\rho,\Phi)<0.2$, as these have circular orbits whose motion is predominantly perpendicular to the line of sight; the points in these regions would have their velocity errors greatly amplified if included, and would confuse the following presentation.  

In Fig.\ 1, we compare the spatial distribution of the \ion{H}{1} with the corresponding distribution of in-plane circular velocities.  As seen in Fig.\ 1, much of the \ion{H}{1} defines a rotation curve that is flat beyond an in-plane radius of $\sim$3 kpc with $v_{\rm circ}\sim80$ km $s^{-1}$, 20 km $s^{-1}$ faster than that derived from the carbon stars.  Correcting the carbon star rotation amplitude for asymmetric
drift, as done by vdM02, we find $v_{\rm circ}\sim74$km s$^{-1}$,
approaching the rotation amplitude of the \ion{H}{1}.  However, there are also several \ion{H}{1} components with clearly distinct kinematic signatures: 1) a component that joins the main rotation curve at $R^\prime\sim$3.5 kpc but rises to an apparent in-plane circular velocity of $v_{\rm c,app}\sim200$ km $s^{-1}$ at $R^\prime\sim$5 kpc (labeled ``S''), 2) a component that joins the main rotation curve at $R^\prime\sim$1.5 kpc but whose in-plane circular velocity falls to $v_{\rm c,app}\sim-30$ km $s^{-1}$ at $R^\prime\sim$4.5 kpc (labeled ``E+B''), 3) an offshoot from component 2 with $v_{\rm c,app}\sim-100$ km $s^{-1}$ at $R^\prime\sim$5 kpc (also encompassed by the label ``E+B''), and 4) a component that like component 2 joins the main rotation curve at $R^\prime\sim$1.5 kpc, but has a steeper velocity gradient with $v_{\rm c,app}\sim-100$ km $s^{-1}$ at $R^\prime\sim$2.5 kpc (labeled ``E2'').  In Fig.\ 1 (bottom), we spatially select regions
occupied by the apparent tidal arms E, S, B, and W (as defined by
Staveley-Smith et al.\ 2003).  We see that the kinematically distinct
components 1-3 described above can be explained as associated with
arms S, E, and B, respectively; arm W, on the other hand, has
kinematics that follow that of the main rotation curve.  Isolating kinematic component 4, we see that it is associated with what appears to be a spatial extension of arm E.  For the remainder of this paper, we will thus refer to the kinematically distinct components as ``S'' (component 1 as described above), ``E+B'' (components 2 and 3), and ``E2'' (component 4).

In Fig.\ 2 (top), we compare the kinematics of the \ion{H}{1} gas with that of the carbon stars.  The main carbon star rotation curve clearly has a shape similar to that of the \ion{H}{1}, but with lower velocity.  Either by chance or physical association, some of the carbon stars lie within the boundaries of the components S, E+B, and E2 associated with \ion{H}{1} tidal features.  These carbon stars are color-coded in Fig.\ 2 and their spatial locations compared with that of the \ion{H}{1}.  In the case of E2, almost all of the associated carbon stars are also spatially coincident with the \ion{H}{1} within E2, making a strong case for physical association.  In the case of component S, roughly half of the carbon stars are spatially coincident with the arm S \ion{H}{1} gas, while the other half lie on the opposite side of the LMC.  This is not unexpected if arm S has a tidal origin, since tidal forces would act on opposite sides of the galaxy with the same magnitude.  The tip of arm S also lies in the region where we have previously found evidence for a warp in the LMC disk (Olsen \& Salyk 2002), providing further reason to suspect tidally disturbed kinematics.
In the case of component E+B, the picture is less clear.  Some of the carbon stars in E+B also overlap spatially with the \ion{H}{1} arms E and B, but the overall spatial distribution is fairly uniform compared to the whole carbon star sample.  On balance, however, the evidence points to a significant number of carbon stars associated with tidal features in the LMC.  

The majority of the RSGs, like the carbon stars, follow the shape of the main \ion{H}{1} rotation curve, but with {\em higher} apparent peak velocity (Fig. 2, bottom).  A number of RSGs have kinematics similar to that of the E+B \ion{H}{1} and E2 components, while a few have velocities like that of arm S.  Spatially, the RSGs with velocities similar to that of arm S are distributed broadly across the LMC's rim, suggesting that they simply fell into our selection region because of the higher mean velocity of the RSGs.  The E+B and E2-like RSGs, on the other hand, lie in a clustered region in the southwest of the LMC, nowhere close to the tidal arms E \& B.  We can think of two explanations for these stars.  The first is that the stars' kinematics are disturbed by the complex supergiant \ion{H}{1} shell SGS4 (Kim et al.\ 1999), which has an expansion velocity of $\sim$23 km $s^{-1}$ as calculated from the approaching side.  The stars also lie in the vicinity of the LMC's warp, which could be reponsible for their peculiar kinematics.

Excluding the stars located within SGS4 and those with $g(\rho,\Phi)<0.2$, we followed the fitting procedure used for the carbon stars, finding $v_{\rm sys}=265.8$ km s$^{-1}$, $\Theta=145\fdg3$, and $V_0=107.5$ km s$^{-1}$ with a velocity dispersion $\sigma=9$ km s$^{-1}$ around the fit, where we left $\alpha_{\rm CM}$, $\delta_{\rm CM}$, d$i$/d$t$, $R_0/d_0$, and $\eta$ at the values determined for the carbon stars (we found that we were not able to constrain these parameters using the RSGs alone).  Thus, a puzzle presented by the RSGs is why their apparent rotation velocity is so large.  

\section{Discussion and Conclusions}
The accurate LMC proper motion measured by Kallivayalil et al.\ (2006) has provided an opportunity for an improved examination of the kinematics of the LMC's carbon stars and red supergiants, which probe intermediate-age ($\gtrsim$100 Myr, with 5-6 Gyr being typical for LMC carbon stars; Cioni et al.\ 2006) and young ($<$20 Myr) stellar populations, respectively.  We have found that $\sim$7\% of the combined Kunkel et al.\ (1997) and Hardy et al.\ (2006) carbon star sample (43 of 642 stars, excluding the 215 stars whose position angles give them little handle on the LMC's internal kinematics) are both spatially and kinematically associated with previously identified \ion{H}{1} tidal features (Staveley-Smith et al.\ 2003), while an additional $\sim$8\% have kinematics like that of the tidal arms.  

Staveley-Smith et al.\ (2003) asked how the near-infrared stellar distribution in the outer parts of the LMC can be so smooth while the \ion{H}{1} is so clearly tidally disturbed.  Our finding that a significant number of carbon stars are associated with tidal arms that are likely $\lesssim2$ Gyr old (Gardiner \& Noguchi 1996, Mastropietro et al.\ 2005) supports the answer given by Weinberg (2000) that the LMC's extended stellar disk is the product of tidal heating.  Additional support for this interpretation is provided by the fact that on scales of $\sim$100 pc, the line-of-sight velocity dispersion of our RSG sample is $\sim$8 km s$^{-1}$, similar to the standard deviation of the velocity residuals with respect to our kinematic solution fit explicitly to the RSGs.  Adopting instead the kinematic solution fit to the carbon stars, the RSG velocity residuals have a standard deviation about the mean of 17 km s$^{-1}$, similar to the velocity dispersion of the carbon stars themselves.  The implication is that once the young stellar population becomes well mixed dynamically, it will acquire a velocity dispersion of at least 17 km s$^{-1}$.

Given the very different rotation curve velocities found for the carbon stars, \ion{H}{1}, and RSGs, which population best reflects the gravitational potential of the LMC?  The non-uniform sampling of the
carbon stars and the evidence for tidal stripping makes it likely that
their kinematics are affected by non-circular streamlines, such that
their $\sim$74 km s$^{-1}$ circular velocity underestimates the true
mass of the LMC.
If the dynamical center derived from the carbon stars is correct, then the \ion{H}{1} that defines the main rotation curve (Fig. 1) has a very asymmetric distribution about the dynamical center, also making the assumption of circular orbits of uncertain applicability and the \ion{H}{1} rotation speed an underestimate.  Although the RSGs suffer from the same problem in addition to the uncertain dynamical influence of the LMC's supergiant shells, the high rotation speed of 107 km s$^{-1}$ derived for them thus may reflect the underlying mass better than the other tracers.  Unravelling the LMC's complex dynamics is, however, likely to require a much larger spectroscopic survey of the LMC's stellar populations than is currently available.  Finally, we note that throughout this paper, we
have used the carbon star solution as the basis for comparison of the
different kinematic tracers.  Given the evidence for tidal
stripping, this solution may not correctly describe the LMC's gravitational potential.  
We note, however, that because our conclusions are based solely on the differences between kinematic tracers, they are not likely to be sensitive to the exact nature of the LMC's potential.

\acknowledgements
We thank Deidre Hunter and Bill Kunkel for interesting dicussions and
readings of an early version of this paper, and Nick Suntzeff for
providing an electronic copy of carbon star velocities.  We thank the
referee, Lister Staveley-Smith, for a careful reading of the paper and
helpful suggestions for improving it.

\clearpage

\begin{figure}
\epsscale{0.75}
\plotone{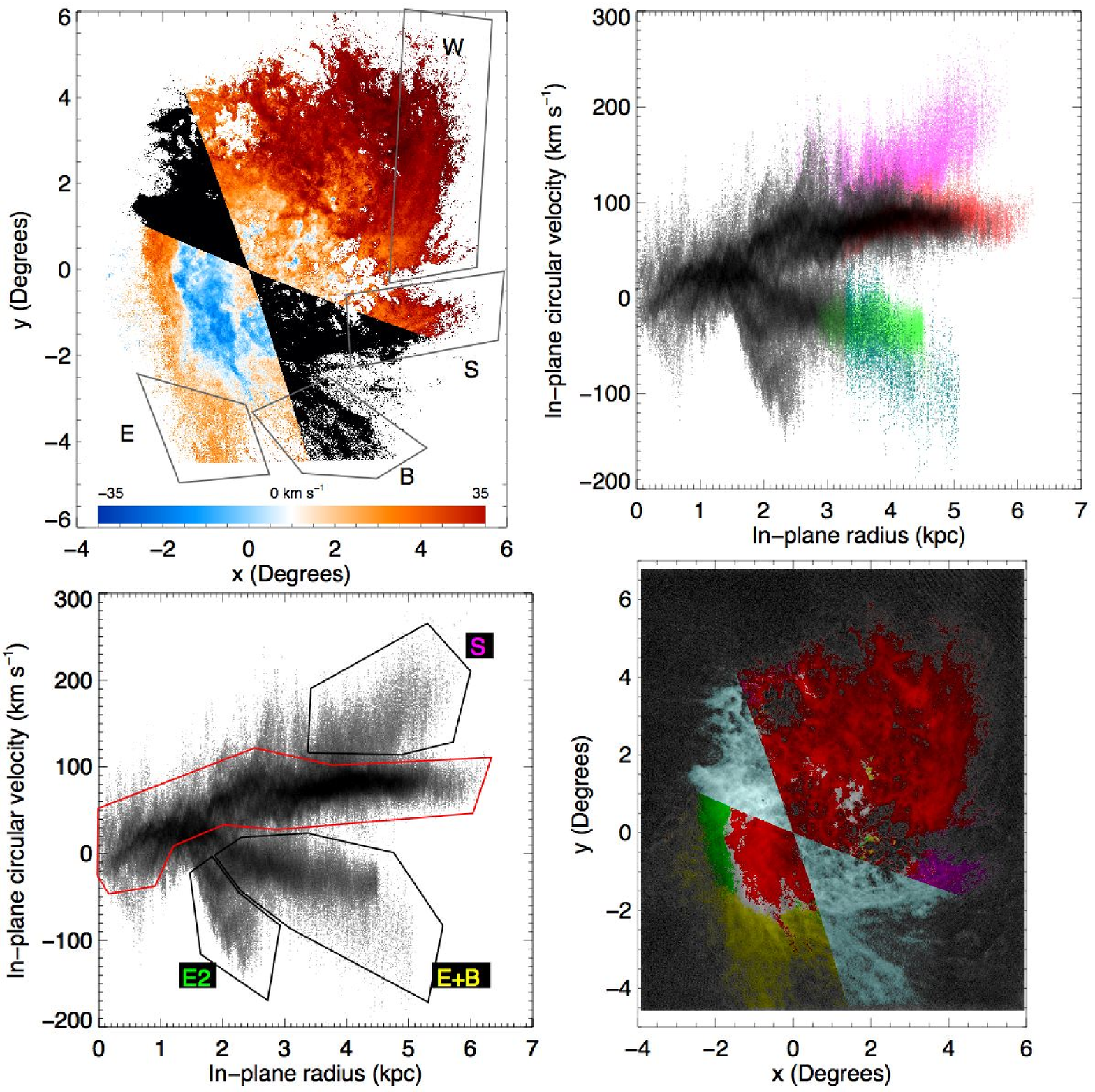}
\caption{Kinematic signatures of the LMC \ion{H}{1} arms identified by Staveley-Smith et al.\ (2003).  {\it Top left:} The emission-weighted velocities from the ATCA+Parkes \ion{H}{1} survey (Kim et al.\ 2003), minus the contribution from the LMC's space motion (Kallavayalil et al.\ 2006), are shown in color along with regions marking the locations of the \ion{H}{1} arms.  The black bowtie-shaped region was excluded from the analysis. The velocity scale runs from -35$\le v \le$35 km s$^{-1}$, while north is up and east to the left.  The origin is fixed to the dynamical center derived from the carbon stars (vdM02), while the image is projected onto the tangent plane.  {\it Top right:} The velocities shown at left have been converted to in-plane circular velocities, as described in the text, and are plotted versus in-plane radius.  Data points belonging in each of the boxes at left are shown in different colors as follows: arm W in red, arm S in magenta, arm B in cyan, and arm E in green. {\it Bottom left:} We identified four regions with distinct kinematic signatures. The red box outlines the main LMC rotation curve and encompasses arm W.  The others are a region with velocities like those of arm S, a region containing arms E and B, and a region with distinct kinematics that we label ``E2''.  {\it Bottom right:} The \ion{H}{1} gas contained within the regions drawn at bottom left are plotted with different colors as follows: red for the main rotation curve, magenta for the arm S region, yellow for the combined arm E and B regions, and green for the region E2.}
\label{fig1}
\end{figure}


\begin{figure}
\epsscale{0.85}
\plotone{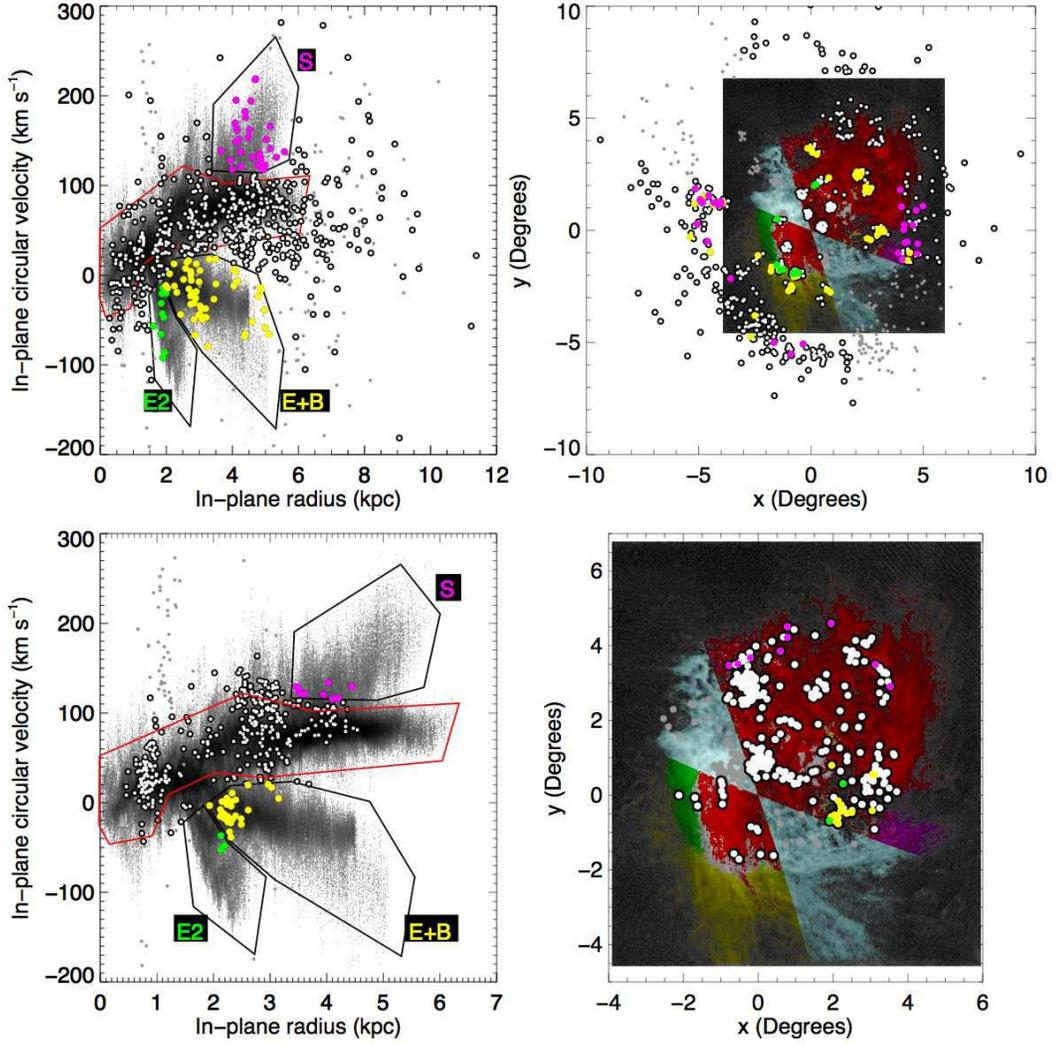}
\caption{Comparing the LMC's stellar kinematics to that of the \ion{H}{1}.  {\it Top left:} As in Fig.\ 1, the line-of-sight \ion{H}{1} velocities converted to in-plane circular velocities are shown vs.\ in-plane radius; the distinct kinematic regions from Fig.\ 1 are also reproduced.  Overplotted as circles are the carbon star velocities.  Carbon stars falling within the regions S (magenta), E+B (yellow), and E2 (green) are labelled in color.  The small grey points are carbon stars falling in the excluded region where in-plane circular orbits are nearly perpendicular to the line of sight.  {\it Top right:}  The positions of the carbon stars and \ion{H}{1} on the sky are shown, with symbols and color labels for the carbon stars as on the left and as in Fig.\ 1 for the \ion{H}{1}.  Many of the carbon stars that have kinematics like that of the tidal \ion{H}{1} arms are also spatially coincident with those arms, implying physical association.  {\it Bottom panels:} Comparing the red supergiant kinematics to that of the \ion{H}{1}. Symbols and colors are as in top panels. The RSGs show apparently faster rotation than the \ion{H}{1}.}
\label{fig2}
\end{figure}


\end{document}